\documentclass[useAMS,usenatbib,usegraphicx]{mn2e}
\usepackage{txfonts}
\DeclareMathAlphabet{\mathpzc}{OT1}{pzc}{m}{it}

\input epsf

\newcommand{\beq}{\begin{equation}}
\newcommand{\eeq}{\end{equation}}

\newcommand{\bneq}{\begin{eqnarray}}
\newcommand{\eneq}{\end{eqnarray}}
\newcommand{\bet}{\begin{table}}
\newcommand{\et}{\end{table}}
\newcommand{\btab}{\begin{tabular}}
\newcommand{\etab}{\end{tabular}}

\title[Kinematic sub-populations in dwarf spheroidal galaxies]{Kinematic sub-populations in dwarf spheroidal galaxies}

\author[U. Ural, et al.]
{U\v gur Ural$^1$, Mark I. Wilkinson$^1$, 
Andreas Koch$^2$, 
Gerard Gilmore$^3$,\newauthor 
Timothy C. Beers$^4$, 
Vasily Belokurov$^3$, 
N. Wyn Evans$^3$, 
Eva K. Grebel$^5$,\newauthor 
Simon Vidrih$^5$\thanks{Humboldt Research Fellow}, 
Daniel B. Zucker$^3$\\\\
$^1$Department of Physics \& Astronomy, University of Leicester, University Road, Leicester, LE1 7RH, United Kingdom\\
$^2$UCLA, Department of Physics and Astronomy,430 Portola Plaza, Los Angeles, CA 90095-1547, USA\\
$^3$Institute of Astronomy, University of Cambridge, Madingley Road, Cambridge, CB3 OHA, UK\\
$^4$Department of Physics \& Astronomy, CSCE: Center for the Study of Cosmic Evolution, and JINA: Joint Institute for Nuclear Astrophysics,\\ Michigan State University, East Lansing, MI 48824, USA\\
$^5$Astronomisches Rechen-Institut, Zentrum f\"{u}r Astronomie der Universit\"{a}t Heidelberg, M\"{o}nchhofstr. 12-14, D-69120  
Heidelberg, Germany\\
}
\begin{document} 

\maketitle
\begin{abstract} 
We present new spectroscopic data for twenty six stars in the
recently-discovered Canes Venatici~I (CVnI) dwarf spheroidal
galaxy. We use these data to investigate the recent claim of the
presence of two dynamically inconsistent stellar populations in this
system~\citep{Ibata2006}. We do not find evidence for kinematically
distinct populations in our sample and we are able to obtain a mass
estimate for CVnI that is consistent with all available data,
including previously published data. We discuss possible differences
between our sample and the earlier data set and study the general
detectability of sub-populations in small kinematic samples. We
conclude that in the absence of supporting observational evidence (for
example, metallicity gradients), sub-populations in small kinematic
samples (typically fewer than 100 stars) should be treated with
extreme caution, as their detection depends on multiple parameters and
rarely produces a signal at the 3$\sigma$ confidence level. It is
therefore essential to determine explicitly the statistical
significance of any suggested sub-population.
\end{abstract}


\section{Introduction}
\label{sec:intro}

It is now widely accepted that the dwarf spheroidal (dSph) satellite
galaxies of the Milky Way and Andromeda are the most dark matter
dominated stellar systems known in the
Universe~\citep[e.g.][]{Mateo1998}. Over the past two decades, a
significant amount of observational work has focussed on quantifying
both the amount of dark matter in these systems, and its spatial
distribution~\citep[e.g.][]{Gilmore2007,Walker2007}.  Although recent
numerical simulations have shown that many of the dSphs may not be
immune to tidal disturbance by the Milky
Way~\citep[e.g.][]{Munoz2008,Lokas2008}, their observed properties
still require the presence of massive dark matter haloes which protect
them against complete tidal disruption. The dSphs thus provide us with
nearby laboratories in which to test dark matter theories.

Given that dSphs occupy the low luminosity end of the galaxy
luminosity function, their star formation histories provide useful
insights into the star formation process. Analyses of spatial
variations in colour-magnitude diagram morphology provided early
evidence of population gradients in a number of
dSphs~\citep[e.g.][]{Harbeck2001}. More recently, evidence of
metallicity gradients has been found using spectroscopic estimates of
[Fe/H]~\citep[e.g.][]{Tolstoy2004,Koch2006,Battaglia2006}. In at least
one case, that of the Sculptor dSph, the metal-rich and metal-poor
populations have significantly different spatial distributions and
kinematics~\citep{Tolstoy2004,Battaglia2008}. Although little evidence
of similar features has been found in other
dSphs~\citep[e.g.][]{Koch2006, Koch2007a, Koch2007b}, the presence of
dynamically distinct stellar populations within dSphs, as well as the
complex interplay between the dynamical, spatial and chemical
properties of their stars, is of great interest as it has implications
for star formation and galaxy evolution.

It is, however, important to note that although the hierarchical
build-up of structure in the standard $\Lambda$-Cold Dark Matter
($\rm{\Lambda\,CDM}$) paradigm implies that satellite galaxies
contribute significantly to the stellar haloes of their hosts,
detailed abundance studies of stars in the more luminous dSphs have
demonstrated that their properties are significantly different from
those of the Milky Way
halo~\citep[e.g.][]{Shetrone2001,Helmi2006}. Among the significant
differences between the halo and the dSphs, the more important
chemical differences are in the
alpha-elements~\citep{Unavane1996,Venn2004}. The observed gradients in
the heavy element distributions are reproduced by the models of
supernova feedback in dSphs developed by~\cite{Marcolini2008}. Thus, it
appears that the primordial dwarf satellites, which were disrupted to
form the Milky Way halo, had stellar populations distinct from those
seen in the present-day dSphs~\citep{Robertson2005, Font2006}.

Given their high estimated mass-to-light ratios, the observed dSphs
are usually identified with the large population of sub-haloes which
are observed to surround Milky Way-sized haloes in cosmological
simulations assuming a standard $\rm{\Lambda\,CDM}$ universe. However,
it was noted early on that the number of dSphs around the Milky Way
was much lower than the expected number of satellite dark matter
haloes~\citep[e.g.][]{Moore1999}. A number of possible explanations
for the apparent lack of Milky Way satellites have been presented in
the literature~\citep[e.g.][]{Stoehr2002, Diemand2005, Moore2006,
Strigari2007, Simon2007, Bovill2008}. All these models are based on
the reasonable postulate that out of the full population of
substructures around the Milky Way, the observed dSphs are merely the
particular subset which (for reasons of mass, orbit, formation epoch,
re-ionisation, etc.) were able to capture gas, form stars and survive
any subsequent tidal interactions with the Milky Way.

In addition, the ratio between the predicted and observed
numbers of dwarf galaxies has decreased significantly in the past few years 
due to the
discovery of nine new Milky Way dSph
satellites~\citep{Willman2005,Zucker2006a,Zucker2006b,Belokurov2006,Belokurov2007,Walsh2007}
in the data from the Sloan Digital Sky
Survey~\citep[SDSS;][]{York2000}.  Since the SDSS covers only about
one fifth of the sky, it is thus likely that the total number of
satellites surrounding the Milky Way may be at least a factor of five
larger than previously thought, although the extrapolation from the
SDSS survey to the whole sky requires careful analysis~\citep[see
e.g.][]{Tollerud2008} . In order to compare the properties of the newly
discovered satellites with those of sub-haloes in cosmological
simulations, as well as to confirm their nature as true satellite
galaxies of the Milky Way, as opposed to star clusters or disrupted
remnants, spectroscopic observations of their member stars are
essential in order to estimate dynamical masses from the observed
stellar kinematics. The extremely low luminosities of these objects
~\citep[in some cases as low as $10^3$L$_\odot$: ][]{Martin2008b},
present significant observational challenges as the kinematic data
sets are small, making it difficult to obtain statistically
significant results.

The Canes Venatici~I (CVnI) dSph is the brightest of the newly
discovered population of very faint SDSS
dSphs~\citep{Zucker2006a}. ~\cite{Ibata2006} presented spectra for a
sample of CVnI member stars obtained using the DEIMOS spectrograph
mounted on the Keck telescope. They identified two kinematically
distinct stellar populations in this data set: an extended metal-poor
population with high velocity dispersion and a centrally-concentrated
metal-rich population with a dispersion of almost zero. Their analysis
of the mass of CVnI suggested that the two populations might not be in
equilibrium as the mass profiles obtained based on the individual
populations were inconsistent with each other.  However, a subsequent
study of CVnI by ~\cite{Simon2007}, using a larger sample of Keck
spectra, did not reproduce this bimodality.

An important outstanding question is whether the ultra-faint dSphs
represent the low-luminosity tail of the dSph population, or are
instead the brightest members of a population of hitherto unknown
faint stellar systems, distinct from both dSphs and star clusters.
The presence of multiple, distinct kinematic populations in a
low-luminosity dSph would set it apart from the majority of
low-luminosity star clusters. In addition, the presence of a spread in
the stellar abundances would suggest an association with the brighter
dwarf galaxies and would also be interesting in terms of its
implications for star formation. It is thus important to determine
whether the sub-population identified by ~\cite{Ibata2006} in CVnI is
real. One goal of our study was to shed some light on this issue by
using spectra obtained with a different spectrograph to those in the
previous two studies of CVnI. In addition, we wanted to investigate
the extent to which sub-populations can be reliably detected in the
very small kinematic data sets which are observable for the
ultra-faint dSphs.

In addition to their potential importance for probing the star
formation histories of dSphs, kinematic substructures can be used to
test another key feature of the hierarchical structure formation
paradigm. The fact that dark matter clustering occurs on all scales
means that the dSph satellites of the Milky Way are likely to be in
the process of accreting their own population of smaller
satellites. Although these substructures may not have been able to
form their own stars, they may be able to acquire stars from their
host dSph. They would then be detectable as localised populations with
mean velocity and/or velocity dispersion distinct from that of the
dSph. Populations with these properties have, in fact, been detected
in the Ursa Minor and Sextans
dSphs~\citep{Kleyna2003,Walker2006}. Once a dSph halo begins to fall
into the Milky Way, it will cease to accrete new satellites as any
nearby substructures will rapidly be removed by the tidal field of the
Milky Way and the high relative velocities in the Milky Way halo will
preclude the capture of new satellites. Due to the short internal
dynamical timescales in dSphs (typically a few hundred Myr), any
remaining internal substructures will subsequently be destroyed on
timescales of at most a few Gyrs if dSph haloes are cusped, although
they can survive much longer if their haloes are
cored~\citep{Kleyna2003}. In the standard cusped-halo picture, only
those satellites which have been interacting with the Milky Way for
less than a few internal dynamical times, either because they are
currently passing the Milky Way for the first time as may be the case
for the Leo~I dSph~\citep[]{Mateo2008} or the Magellanic Clouds
~\citep[]{Kallivayalil2006, Besla2007,Piatek2008} or because their
crossing times are larger~\citep[e.g. the Magellanic
Clouds:][]{vanderMarel2002}, would be expected to exhibit localised
kinematic substructure. If localised substructures were found to be
common in dSphs, this could be difficult to reconcile with a picture
in which dSphs occupy cusped haloes. Given that the level of
substructure above a given mass fraction is a function of halo mass
\citep{Gao2004}, the expected numbers of sub-haloes per dSph requires
further investigation by means of cosmological simulations. However,
the importance of comparing the level of substructure in dSphs with
the results of numerical simulations adds further motivation to our
goal of establishing the level of confidence with which
sub-populations can be detected in small data sets.

The outline of the paper is as follows. In Section~\ref{sec:cvn}, we
present a new kinematic data set for stars in CVnI, based on spectra
obtained with the Gemini telescope, and calculate a mass estimate for
the galaxy from these data. In Section~\ref{sec:pop}, we look for
kinematic sub-populations in our data, and compare our findings with
those of ~\cite{Ibata2006}. Section~\ref{sec:det} discusses the
general detectability of sub-populations in small kinematic data sets.
Finally, in Section~\ref{sec:conc} we draw some general conclusions
and suggest possible differences between the two data sets for CVnI
that we have compared.

\section{Canes Venatici I}
\label{sec:cvn}

\subsection{Data Reduction}
\label{sec:reduction}

\begin{figure}
  \includegraphics[width=0.5\textwidth]{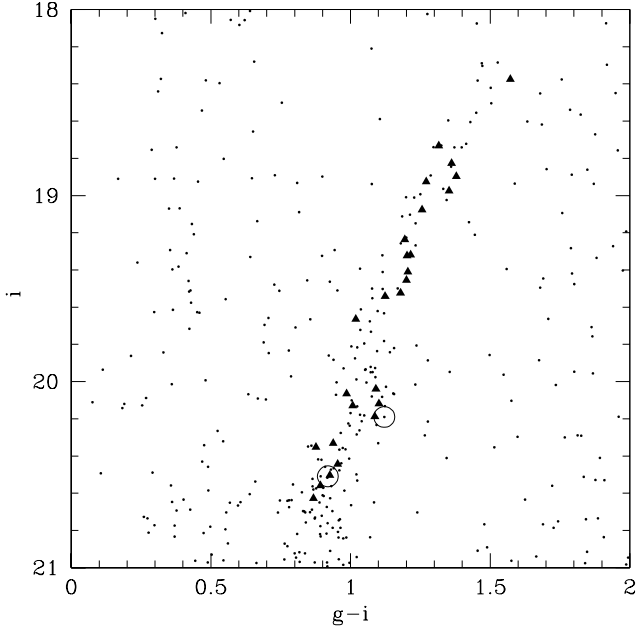}
\caption{SDSS ($g-i,i$) color-magnitude diagram for stars in a field of radius 
15 arcmin centred on CVnI. Our likely CVnI members are indicated as
solid triangles. Two velocity outliers are shown as open circles.}
\label{fig:fcmd1}
\end{figure}

\begin{figure}
  \includegraphics[width=0.6\textwidth]{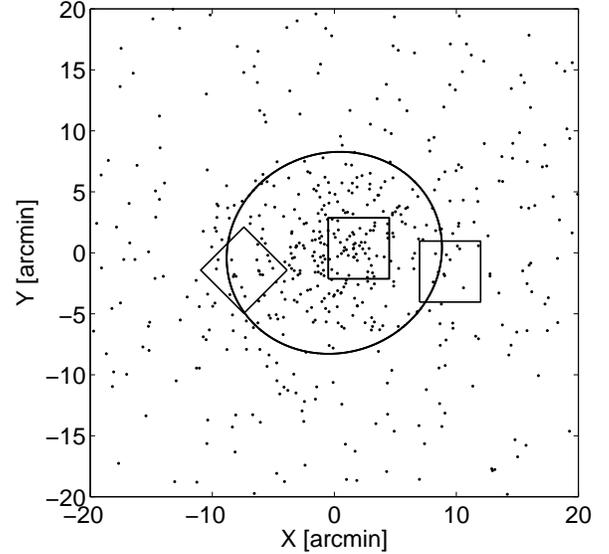}
\caption{Distribution of our GMOS target fields in CVnI. The data 
points show the positions of stars satisfying our CMD selection
cut. The slight excess of stars in the region $-10 < X < 10$ and $-5 <
Y < 5$ indicates the location of the main body of CVnI. The ellipse
shows the half-light radius of the system, with semi-major axis $8.9$
arcmin
\protect\citep{Martin2008b}.}
\label{fig:fields}
\end{figure}

Twenty eight stars in the CVnI dSph were observed on 2007 March 26 and
2007 April 7 and 8 using the GMOS-N spectrograph mounted on the Gemini
North telescope. Our targets were chosen by cross-matching of GMOS-N
pre-images (taken in the i-band) with existing SDSS photometry. As
Figure~\ref{fig:fcmd1} shows, all the selected stars lie in the red
giant branch (RGB) region of the CVnI colour-magnitude diagram
(CMD). A total of three GMOS slit-masks were observed, with the
spectra centred on the spectral region containing the Ca triplet
region (around $860$nm). Our masks covered three distinct fields in
CVnI. Figure~\ref{fig:fields} shows the locations of the fields
relative to the spatial distribution of stars in CVnI. The masks were
cut with slitlets of width 0.75 arcsec.

The GMOS detector consists of three adjacent CCDs. As the dispersion
axis of the slits is perpendicular to the spaces between the CCDs, the
spectra contain gaps corresponding to the inter-CCD gaps. In order to
achieve continuous wavelength coverage throughout the spectral region
of interest, each mask was observed in two configurations with
different central wavelengths ($855$nm and $860$nm). All observations
were taken using the R831+\_G5302 grating and CaT\_G0309 filter, with
2$\times$4 spectral and spatial binning, respectively. The spectra
thus obtained have a nominal resolution of 3600. The three fields were
observed for a total of 10,800s, 9,000s and 12,600s, respectively,
with the observations divided into individual exposures of 1800s to
facilitate cosmic ray removal.

\begin{figure}
  \includegraphics[width=0.5\textwidth]{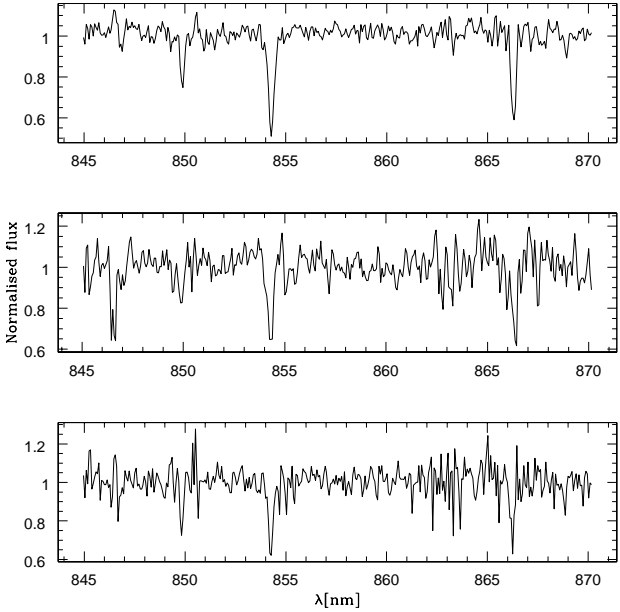}
\caption{Three sample spectra for stars of magnitude i$=19.7$ (top), 
i$=20.6$ (middle) and i$=20.1$ (bottom). The quality of the spectrum in the bottom
 panel is typical of the majority of our stars.}
\label{fig:spectrum}
\end{figure}

The raw data were reduced using the standard {\tt gemini} reduction
package which is run within the Image Reduction and Analysis Facility
(IRAF)\footnote{IRAF is distributed by the National Optical Astronomy
Observatories, which are operated by the Association of Universities
for Research in Astronomy Inc. (AURA), under cooperative agreement
with the National Science Foundation.} environment. All data were
first bias subtracted and flat-field corrected. The individual
spectral traces were identified from flat field images (obtained using
Quartz halogen continuum lamp exposures).  The wavelength calibration
of the spectra was performed using CuAr lamp exposures adjacent in
time to the science exposures as calibration frames. The typical
r.m.s.  uncertainty in the wavelength calibration, obtained by fitting
a polynomial to the line positions in the CuAr spectra, was $0.01$\AA,
which corresponds to a velocity error of $\sim0.4$\,km\,s$^{-1}$ at a
wavelength of $860$nm. This wavelength solution was then applied to
the reduced science spectra. Sky subtraction was performed by using
the sky flux in the regions of the slit not dominated by light from
the target to estimate the sky spectrum. Finally, the object spectra
were extracted from the CCD images using a fifth order Chebyshev
polynomial fit. Figure~\ref{fig:spectrum} shows examples of a good
quality spectrum (top panel), a low quality spectrum (middle panel),
and typical quality spectrum (bottom).

\subsection{Velocities}
The velocities of the stars were calculated using the {\tt fxcor} task
in IRAF to cross-correlate the stellar spectral lines with the lines
in a template Ca triplet spectrum. The synthetic template consisted of
three Gaussian lines at the wavelengths of the Ca triplet lines, whose
widths were chosen to match those typical of RGB stars. We first
cross-correlated the individual science exposures as a preliminary
diagnostic of whether any spectra were obviously anomalous and should
be excluded. As none of the spectra seemed to have serious problems,
all frames were used in the velocity calculations and we combined all
heliocentric-corrected exposures of the same mask together in order to
increase the signal-to-noise.

\begin{figure}
  \includegraphics[width=0.5\textwidth]{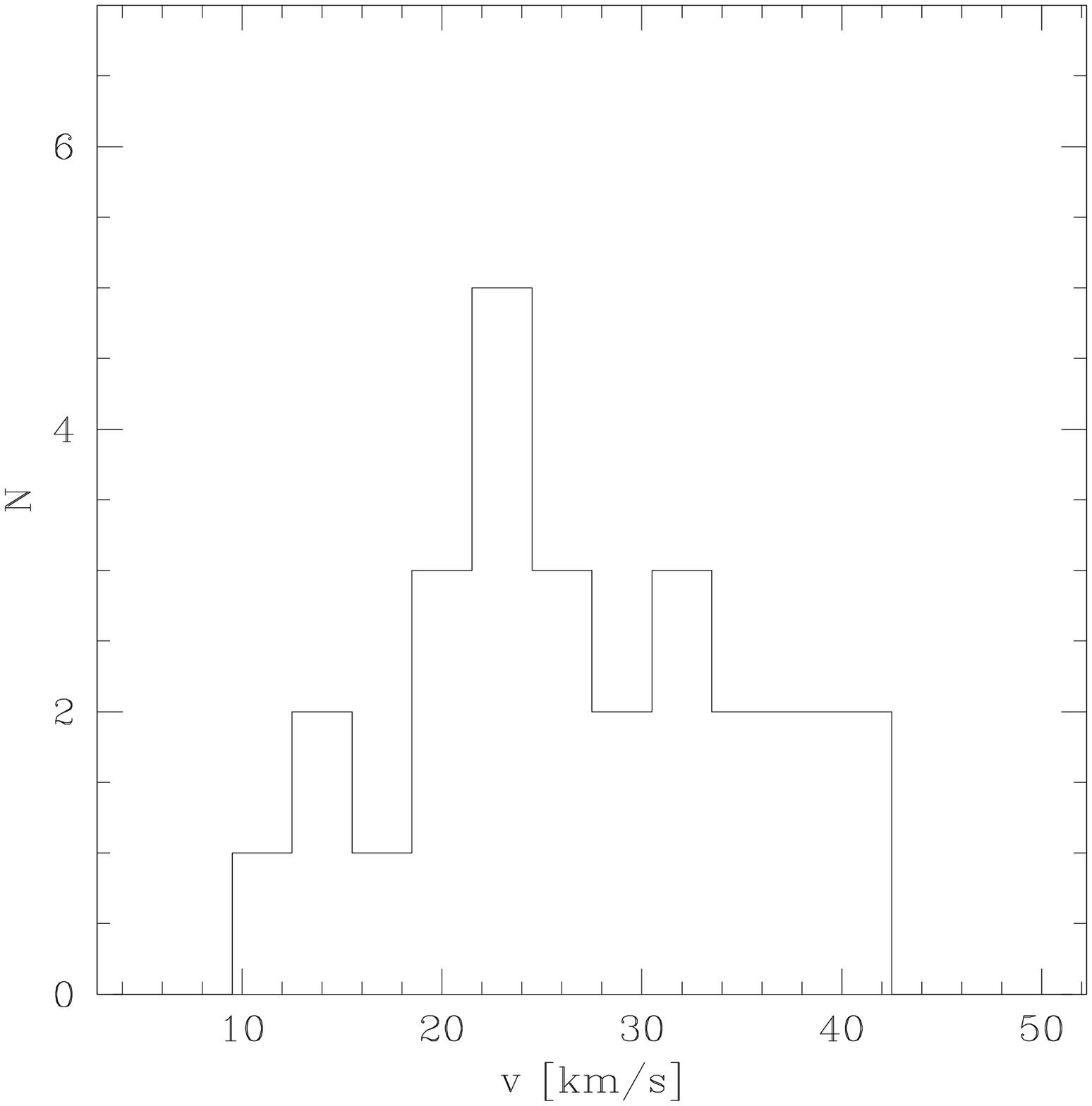}
\caption{Velocity histogram for our data set of 26 likely members of CVnI. 
Two obvious velocity outliers (at $44$\,km\,s$^{-1}$ and
$-39.1$\,km\,s$^{-1}$) have been excluded in from the figure.}
\label{fig:fv1}
\end{figure}

\begin{figure}
  \includegraphics[width=0.5\textwidth]{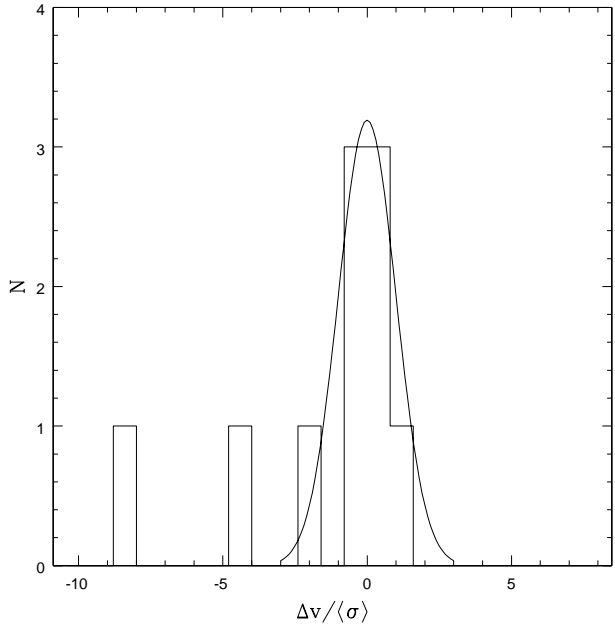}
\caption{Histogram of the normalised differences in estimated 
velocity for the ten stars observed both by us and
\protect\cite{Ibata2006}. Differences have been normalised by the
combined error from the two estimates (see text for details). Apart
from the two significant outliers, the distribution is close to the
overplotted Gaussian.}
\label{fig:fv2}
\end{figure}

The {\tt fxcor} task returns estimated velocity uncertainties which
are based on the Tonry-Davis Ratio for the fitted cross-correlation
peak. These errors are often found not to be an accurate reflection of
the true uncertainties~\citep[see e.g.][]{Kleyna2002,Munoz2005}. In
order to estimate the actual uncertainty in our velocity
determinations, we measured separately the velocities $v_1$ and $v_2$
for the spectra with central wavelengths 855nm and 860nm,
respectively. We combined these estimates to obtain the mean velocity
for each star $\overline{v} = 0.5(v_1 + v_2)$ and defined a $\chi^2$
statistic via
\beq
\chi^{2}=\frac{(v_{1}-\overline{v})^{2}}{(dv_{1})^{2}}+\frac{(v_{2}-\overline{v})^{2}}{(dv_{2})^{2}} ,
\label{eq:echisq}
\eeq
where $dv_1$ and $dv_2$ are the formal errors returned by {\tt
fxcor}. We then rescaled the velocity errors in our sample by a factor
$f$ so that the sum of equation~\ref{eq:echisq} over all stars was
$2N$, where $N$ is the size of the velocity sample. Finally, using the
rescaled errors, we calculated $P(\chi^2)$ for each star, using the
routine {\tt gammq} from Numerical Recipes~\citep{Press1991}. The
final velocities and errors are given in
Table~\ref{tab:tres}. Following the error rescaling, only one star was
found to have an extremely low value of $P(\chi^2)$ ($<10^{-4}$). As
Table~\ref{tab:tres} shows, this star also has the largest velocity in
the sample and a relatively large estimated velocity error, possibly
due its low signal-to-noise ratio, and we therefore excluded it from
our final sample. We also excluded one star which has very different
radial velocity $v_R=-39.1$\,km\,s$^{-1}$ compared to mean velocity of
the rest of our target stars ($25.8\pm 0.3$\,km\,s$^{-1}$; see
Section~\ref{sec:mass}). Figure~\ref{fig:fv1} shows the velocity
histogram for our final sample consisting of $26$ stars. We note that
our sample includes 10 stars from the \cite{Ibata2006} and
\cite{Martin2007} sample and Figure~\ref{fig:fv2} is a histogram
representing the difference between the velocities of these stars from
both studies in terms of their $1\sigma$ measurement
uncertainties. Thus, we calculate $\Delta \rm v /
\langle\sigma\rangle$ as
$(v_{Keck}-v_{GMOS})/\sqrt{dv^{2}_{Keck}+dv^{2}_{GMOS}}$, after
applying a velocity shift of -3.4 km\,s$^{-1}$ to our estimates in
order to bring the median of the two data sets together. The plot
shows that apart from the two outliers, at 8.6 and 4.6$\sigma$, with
very different velocities in the two sets, the differences are
normally distributed. The outliers are possibly stars in binary
systems which have changed their velocity between the two
observations. The
\cite{Ibata2006} data were taken in May 2006, i.e. around ten months
earlier than our data. The observed velocity differences of
$8-10$km\,s$^{-1}$ over this baseline are consistent with tight binary
orbits.

\subsection{Metallicities}
\begin{figure}
  \includegraphics[width=0.5\textwidth]{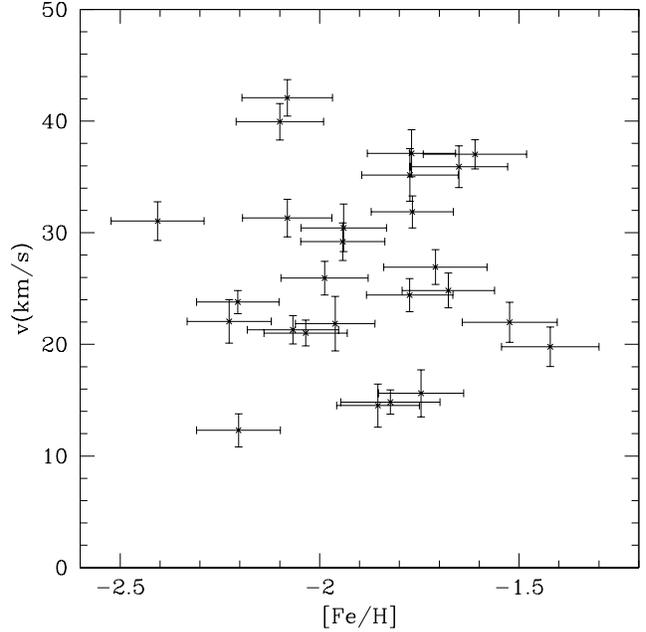}
\caption{Line of sight velocity versus [Fe/H] for our sample of likely CVnI members.
The two obvious velocity outliers have been excluded. }
\label{fig:fres1}
\end{figure}

It is now well-established that the line strength of the near-infrared
Ca triplet lines in the spectrum of an RGB star can be used to
estimate the [Fe/H] of the star~\citep[e.g.][]{Armandroff1988,
Armandroff1991, Carrera2007, Bosler2007}. We note that the accuracy of
this method may be less reliable when extrapolating below
metallicities of $\sim-2.2$ where globular cluster calibrators are
missing~\citep{Koch2008}, although comparisons of high-vs-low
resolution data by \cite{Battaglia2008} have shown that CaT-based
estimates may be correct down to [Fe/H] $\sim -3$. In practice, we
normalized the spectra using a seventh order Legendre polynomial,
fitted each of the triplet lines using a Penny
function~\citep[see][]{Cole2004}, and integrated the profile over the
standard band passes of~\cite{Armandroff1988}. The final [Fe/H]
metallicities, on the scale of~\cite{Carretta1997}, were calculated
using the calibration of ~\citeauthor{Rutledge1997a}
(\citeyear{Rutledge1997a,Rutledge1997b}), namely
\beq
{\rm[Fe/H]}=-2.66+0.42[\Sigma W+0.64({\rm V}-{\rm V}_{\rm HB})] ,
\label{eq:eeqcalib}
\eeq
where we parameterised the line strength of the Ca triplet as
\beq
\Sigma W=0.5*w_{1}+w_{2}+0.6*w_{3} ,
\label{eq:eeqwidth}
\eeq
where $w_{1}$, $w_{2}$ and $w_{3}$ are the widths of the individual
lines.  In Eq.~\ref{eq:eeqcalib}, V is the V-band magnitude of the
star, and V$_{\rm HB}$ is the magnitude of the horizontal branch of
the system. For the latter, we used a value of $\rm V_{HB}=22.4$,
obtained by visual inspection of the (V-I,V) colour-magnitude diagram
of CVnI. We note that this is very similar to the value of $\rm V_{HB}=22.5$ used by
\cite{Martin2008a}. The uncertainty of $\sim 0.1$ magnitudes in V$_{\rm HB}$
gives rise to a negligible additional uncertainty in our [Fe/H]
estimates. The random errors on the [Fe/H] metallicities were
calculated using the formalism of \cite{Cayrel1988} for the errors on
the single line widths and are based on the spectral signal-to-noise
ratio. These were then propagated through the calibration equations,
accounting for photometric errors.  The final metallicity estimates
are given in Table~\ref{tab:tres}. Figure~\ref{fig:fres1} shows the
distribution of velocity versus [Fe/H] for our CVnI sample. The
error-weighted mean [Fe/H] is $-1.9\pm 0.02$ compared to the value of
$-2.09\pm 0.02$ found by ~\cite{Simon2007}. We note that all previous
studies of CVnI have found a significant spread in [Fe/H], of order
0.5 dex~\citep{Ibata2006,Simon2007,Kirby2008}, and our value thus lies
within the range of previous estimates.  As the figure shows, there
appear to be no obvious correlations between velocity and [Fe/H] in
our sample.

\begin{table*}
\begin{center}
\begin{tabular} {l l c c c c c c c c c r}
\hline
$\alpha$ (J2000)  & $\delta$ (J2000) & V & I & g & i&  $v_{\rm r}$ (km\,s$^{-1}$) & $dv_{\rm r}$  & $\sum$W & d$\sum$W & [Fe/H] & $d$[Fe/H] \\\hline
$13\,28\,10.07$ &  $+33\,33\,41.6$ & $ 20.2$ &  $ 19.2 $  & $  20.7 $  & $  19.7$ &  $29.2$ &  $1.7$   &  $  3.12 $   & $  0.07 $    & $  -1.93 $  &   $ 0.10  $ \\
$13\,28\,10.31$ &  $+33\,33\,06.0$  & $ 19.8$ &  $ 18.6 $  & $  20.3 $  & $  19.1$ &  $39.9$ &  $1.6$   &  $  3.02 $   & $  0.05 $    & $  -2.09 $  &   $ 0.11  $ \\ 
$13\,28\,16.78$ &  $+33\,32\,54.0$ & $ 20.6$ &  $ 19.6 $  & $  21.1 $  & $  20.0$ &  $21.9$ &  $2.4$   &  $  2.83 $   & $  0.08 $    & $  -1.95 $  &   $ 0.10  $ \\
$13\,28\,15.00$ &  $+33\,32\,01.6$  & $ 20.8$ &  $ 19.9 $  & $  21.3 $  & $  20.3$ &  $30.4$ &  $2.1$   &  $  2.72 $   & $  0.14 $    & $  -1.93 $  &   $ 0.11  $ \\
$13\,28\,18.31$ &  $+33\,33\,17.3$ & $ 20.8$ &  $ 19.9 $  & $  21.2 $  & $  20.4$ &  $14.5$ &  $1.9$   &  $  2.96 $   & $  0.11 $    & $  -1.84 $  &   $ 0.10  $ \\
$13\,28\,08.49$  &  $+33\,33\,29.4$ & $ 21.2$ &  $ 20.2 $  & $  21.5 $  & $  20.6$ &  $35.2$ &  $2.4$   &  $  2.94 $   & $  0.19 $    & $  -1.76 $  &   $ 0.12  $ \\
$13\,28\,08.82$  &  $+33\,34\,41.2$ & $ 19.9$ &  $ 18.8 $  & $  20.4 $  & $  19.2$ &  $25.9$ &  $1.5$   &  $  3.22 $   & $  0.05 $    & $  -1.97 $  &   $ 0.11  $ \\
$13\,28\,10.35$ &  $+33\,34\,27.7$ & $ 20.2$ &  $ 19.1 $  & $  20.7 $  & $  19.5$ &  $24.4$ &  $1.5$   &  $  3.55 $   & $  0.06 $    & $  -1.76 $  &   $ 0.11  $ \\
$13\,28\,16.23$ &  $+33\,34\,09.9$  & $ 20.2$ &  $ 19.0 $  & $  20.7 $  & $  19.5$ &  $24.8$ &  $1.6$   &  $  3.81 $   & $  0.10 $    & $  -1.66 $  &   $ 0.12  $ \\ 
$13\,28\,11.31$ &  $+33\,35\,06.1$  & $ 20.8$ &  $ 19.7 $  & $  21.3 $  & $  20.2$ &  $21.9$ &  $1.8$   &  $  3.77 $   & $  0.15 $    & $  -1.51 $  &   $ 0.12  $ \\
$13\,28\,07.19$  &  $+33\,36\,05.4$  & $ 20.2$ &  $ 19.1 $  & $  20.7 $  & $  19.5$ &  $23.8$ &  $1.0$   &  $  2.51 $   & $  0.08 $    & $  -2.19 $  &   $ 0.10  $ \\
$13\,28\,13.74$ &  $+33\,35\,55.6$ & $ 19.7$ &  $ 18.5 $  & $  20.2 $  & $  18.9$ &  $21.3$ &  $1.3$   &  $  3.19 $   & $  0.06 $    & $  -2.05 $  &   $ 0.11  $ \\
$13\,28\,24.02$ &  $+33\,35\,32.7$ & $ 21.0$ &  $ 20.1 $  & $  21.4 $  & $  20.5$ &  $44.0$ &  $3.2$   &  $  2.23 $   & $  0.18 $    & $  -2.09 $  &   $ 0.11  $ \\
$13\,27\,38.76$ &  $+33\,32\,55.3$ & $ 20.0$ &  $ 18.8 $  & $  20.5 $  & $  19.3$ &  $12.3$ &  $1.4$   &  $  2.65 $   & $  0.05 $    & $  -2.19 $  &   $ 0.10  $ \\
$13\,27\,33.80$ &  $+33\,32\,57.3$ & $ 20.8$ &  $ 19.7 $  & $  21.2 $  & $  20.1$ &  $31.9$ &  $1.4$   &  $  3.20 $   & $  0.09 $    & $  -1.75 $  &   $ 0.10  $ \\
$13\,27\,28.41$ &  $+33\,33\,26.3$ & $ 20.8$ &  $ 19.7 $  & $  21.3 $  & $  20.2$ & $-39.1$ &  $2.0$   &  $  4.26 $   & $  0.13 $    & $  -1.29 $  &   $ 0.12  $ \\
$13\,27\,28.29$ &  $+33\,32\,10.6$ & $ 19.6$ &  $ 18.3 $  & $  20.2 $  & $  18.8$ &  $26.9$ &  $1.6$   &  $  4.06 $   & $  0.13 $    & $  -1.70 $  &   $ 0.13  $ \\
$13\,27\,30.35$ &  $+33\,32\,04.3$ & $ 21.1$ &  $ 20.1 $  & $  21.5 $  & $  20.6$ &  $15.6$ &  $2.1$   &  $  3.01 $   & $  0.13 $    & $  -1.73 $  &   $ 0.11  $ \\
$13\,27\,30.78$ &  $+33\,29\,38.9$ & $ 20.6$ &  $ 19.6 $  & $  21.1 $  & $  20.1$ &  $19.8$ &  $1.8$   &  $  4.12 $   & $  0.14 $    & $  -1.41 $  &   $ 0.12  $ \\
$13\,27\,31.21$ &  $+33\,29\,59.2$ & $ 19.3$ &  $ 17.9 $  & $  19.9 $  & $  18.4$ &  $14.8$ &  $1.1$   &  $  4.01 $   & $  0.05 $    & $  -1.81 $  &   $ 0.12  $ \\
$13\,27\,16.19$ &  $+33\,32\,22.2$ & $ 20.1$ &  $ 18.9 $  & $  20.6 $  & $  19.4$ &  $21.0$ &  $1.2$   &  $  2.98 $   & $  0.05 $    & $  -2.02 $  &   $ 0.10  $ \\
$13\,28\,52.91$ &  $+33\,29\,26.0$ & $ 19.8$ &  $ 18.5 $  & $  20.3 $  & $  19.0$ &  $31.3$ &  $1.7$   &  $  3.11 $   & $  0.06 $    & $  -2.07 $  &   $ 0.11  $ \\
$13\,28\,46.21$ &  $+33\,30\,49.7$ & $ 21.1$ &  $ 20.1 $  & $  21.4 $  & $  20.5$ &  $35.9$ &  $1.9$   &  $  3.29 $   & $  0.18 $    & $  -1.67 $  &   $ 0.12  $ \\
$13\,28\,50.64$ &  $+33\,31\,20.9$ & $ 20.0$ &  $ 18.8 $  & $  20.5 $  & $  19.3$ &  $42.1$ &  $1.6$   &  $  2.93 $   & $  0.11 $    & $  -2.07 $  &   $ 0.11  $ \\
$13\,28\,50.29$ &  $+33\,32\,31.6$ & $ 19.7$ &  $ 18.4 $  & $  20.3 $  & $  18.9$ &  $37.0$ &  $1.3$   &  $  4.24 $   & $  0.13 $    & $  -1.60 $  &   $ 0.13  $ \\
$13\,28\,45.29$ &  $+33\,31\,35.1$ & $ 21.0$ &  $ 20.0 $  & $  21.4 $  & $  20.4$ &  $37.1$ &  $2.1$   &  $  3.05 $   & $  0.15 $    & $  -1.76 $  &   $ 0.11  $ \\
$13\,28\,53.47$ &  $+33\,33\,22.5$ & $ 20.7$ &  $ 19.7 $  & $  21.1 $  & $  20.1$ &  $22.1$ &  $1.9$   &  $  2.12 $   & $  0.14 $    & $  -2.21 $  &   $ 0.10  $ \\
$13\,28\,44.26$ &  $+33\,34\,11.8$ & $ 19.5$ &  $ 18.3 $  & $  20.0 $  & $  18.7$ &  $31.0$ &  $1.7$   &  $  2.47 $   & $  0.08 $    & $  -2.40 $  &   $ 0.12  $ \\
\hline
\end{tabular}
\caption{Summary of properties of our CVnI data. Columns give: 
(1) Right ascension; (2) Declination; (3,4) V and I magnitudes,
calculated from SDSS photometry using the transformations of Lupton
(2005:
http://www.sdss.org/dr4/algorithms/sdssUBVRITransform.html\#Lupton). Lupton
derived these equations by matching photometry from SDSS Data Release
4 to Stetson's published photometry; (5,6) SDSS g, i magnitudes; (7,8)
radial velocity and error, in km\,s$^{-1}$; (9,10) combined equivalent
width of Ca triplet lines, with error, obtained using
equation~\ref{eq:eeqwidth}; (11,12) estimated metallicity [Fe/H], with
error, obtained using equation~\ref{eq:eeqcalib}. Note that one star
($v = -39.1$\,km\,s$^{-1}$) is a clear outlier from the mean velocity
of $v = 25.8\pm 0.3$\,km\,s$^{-1}$. A second outlier has $v =
44$\,km\,s$^{-1}$ and also a relatively large error of $dv =
3.2$\,km\,s$^{-1}$ for our sample. We therefore exclude these two
stars from our analysis.}
\label{tab:tres}
\end{center}
\end{table*}

\subsection{Mass Calculation}
\label{sec:mass}

In order to estimate the mass of CVnI, we calculate the velocity
dispersion of the system using the new velocity set that we obtained
in the previous section. We use a maximum likelihood method
~\citep[e.g. ][]{Kleyna2004} to calculate the velocity dispersion and
the mean velocity of our data. We apply an iterative $3\sigma$ cut in
velocity - however, we note that this did not remove any stars (ie. it
converged after a single iteration).  Based on the CMD in
Figure~\ref{fig:fcmd1}, we do not expect significant foreground
contamination in our velocity sample, and we therefore use all 26 of
our stars when estimating the dispersion. We find a dispersion of
$\sigma = 7.9^{+1.3}_{-1.1}$\,km\,s$^{-1}$ and a mean velocity of $v=
25.8\pm 0.3$\,km\,s$^{-1}$. The latter is somewhat smaller than the
value of $v =30.9\pm 0.6$\,km\,s$^{-1}$ found by
\cite{Simon2007}. \cite{Ibata2006} found dispersions of
$13.9$\,km\,s$^{-1}$ and $0.5$\,km\,s$^{-1}$ for the two populations
they identified. As we discuss below, we do not find evidence of
multiple populations in our data and we therefore quote only a single
value for the dispersion.

Following ~\cite{Ibata2006} we use our dispersion measurement to
constrain the mass of CVnI. In order to proceed we need to
parameterise the spatial distribution of our data.  We assume that our
tracer population is drawn from a Plummer distribution and we find the
scalelength for which the likelihood of the positions of our tracer
data set is maximised. Based on the positions of our tracer stars
only, we find a Plummer scalelength of $a=4.62$~arcmin, which is
smaller than the value of $8.5\pm 0.5$~arcmin found by
~\cite{Zucker2006a} and $8.9\pm0.4$~arcmin found by ~\cite{Martin2008b}
for the full stellar distribution. The mass is then calculated
using the isotropic Jeans equation ~\citep[][eq. 4.56]{Binney1987},
under the additional assumption of spherical symmetry.

We find a mass of $4.4^{+1.6}_{-1.1} \times10^7 M_\odot$ within the
volume probed by our data (i.e. out to a radius of $11$~arcmin). The
mass-to-light ratio is calculated assuming a luminosity of $\mbox{L}=
(2.3\pm0.3)\times10^5$ L$_\odot$ ~\citep[]{Martin2008b}. Assuming
symmetric errors we find $\mbox{M/L}=192\pm76\,$M$_\odot$/L$_\odot$.
If we take the value of the scalelength reported by \cite{Zucker2006a}
to be that of our tracers, we obtain a mass of
$3.3^{+1}_{-2}\times10^7 M_\odot$. We note that both these estimates
are larger than the mass of M$=(2.7\pm 0.4)\times10^7 M_\odot $
reported by ~\cite{Simon2007} using their larger data set. The
difference is probably due to our assumption of a constant velocity
dispersion profile, while the assumption of mass-follows-light was
implicitly made by those authors. ~\cite{Ibata2006} obtained two very
different mass estimates using the distinct populations which they
identified in their data. An important point to keep in mind while
dealing with small data sets is that the Jeans equations remain valid
for density-weighted averages of the spatial distributions, velocity
dispersion profiles and velocity anisotropy profiles of multiple
tracer populations~\citep[]{Strigari2007}. Thus, it is legitimate to
use a data set which may contain multiple sub-populations when
estimating the mass of the system. As long as all sub-populations
are in dynamical equilibrium, this estimate will be more reliable than
the noisier estimates based on the smaller, individual populations.

\section{Sub-populations}
\label{sec:pop}

\subsection{Canes Venatici I}

As we noted above, ~\cite{Ibata2006} identified two kinematically
distinct populations in CVnI. Given the potential importance of
sub-populations in dSphs discussed in the introduction, we now
investigate whether our data exhibit any evidence of multiple
populations. Although Fig.\ref{fig:fres1} shows a wide scatter in the
abundances that might be due to an extended star formation period, no
clear signature of distinct sub-populations is seen. In order to
confirm this visual impression more quantitatively, we fitted our
velocity distribution with multiple Gaussians and tested the
significance of the fits using Monte Carlo realisations of our
data. Our approach, which is essentially a likelihood ratio test, is
similar to the KMM test ~\citep[]{Ashman1994} which is designed to
detect multiple Gaussian populations with different means and
dispersions within a single data set, although unlike the KMM test we
do not include a determination of which sub-population the individual
stars belong to.

\begin{figure}
  \includegraphics[scale=0.4]{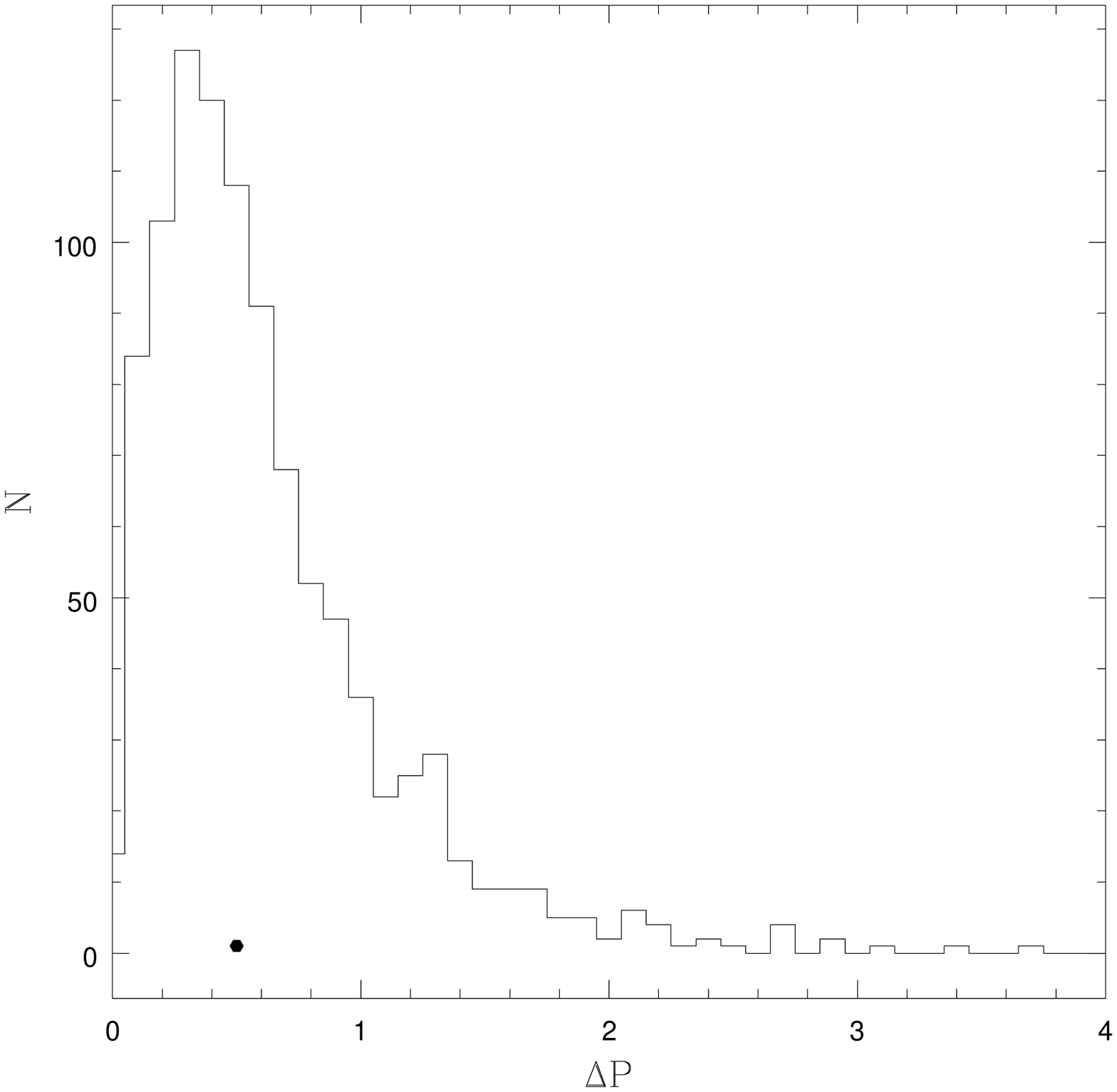}
  \includegraphics[scale=0.4]{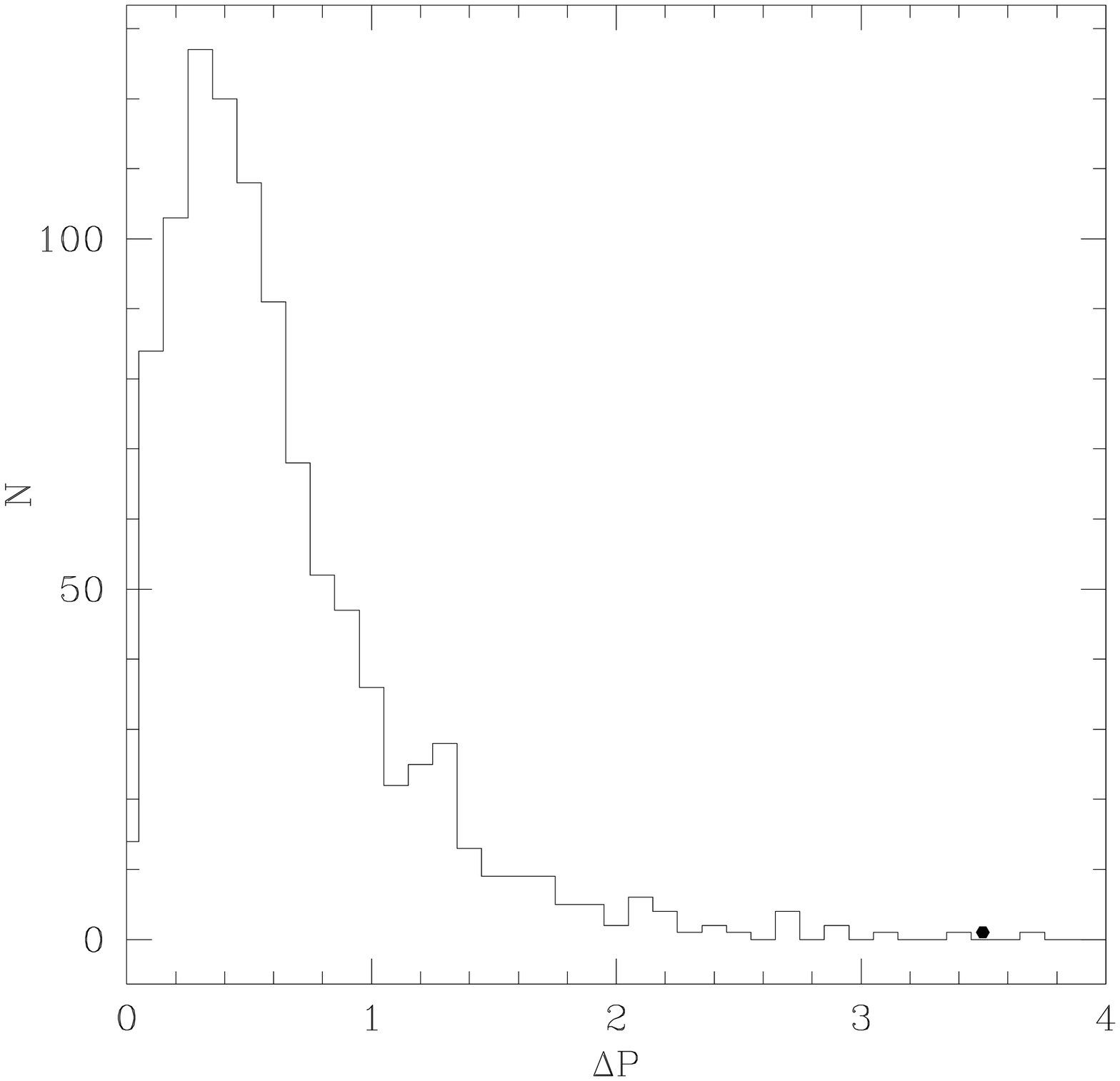}
\caption{Distribution of likelihood ratios $\Delta P = \log P_2-\log P_1$ 
between a single-Gaussian fit ($P_1$) and a two-Gaussian fit ($P_2$)
to 1000 Monte Carlo realisations of 26 stars drawn from a single
Gaussian distribution. The single dots indicate the values obtained
for our GMOS-N data (upper panel) and the Keck data of
\protect\cite{Ibata2006} (bottom panel). Although there is no
evidence of multiple populations in our data, a sub-population
containing a fraction of $40$ per cent of the stars is detected in the
\protect\cite{Ibata2006} data. See text for a detailed discussion.}
\label{fig:frealtest1}
\end{figure}

The first step of this process was to fit a single Gaussian to our
velocity data.  We then repeated the fit for a two-Gaussian model in
which a fraction $f$ of the data belonged to a population with mean
$\overline{v_{1}}$ and dispersion $\sigma_{1}$, and the remaining data
had mean $\overline {v_{2}}$ and dispersion $\sigma_{2}$.  As
expected, the two-Gaussian model yielded higher likelihoods. In order
to determine whether this was only due to the increased number of
fitting parameters or was a real detection, we tested the significance
of the results with artificial data. To do this, we generated 1000
data sets of 26 stars drawn from a single Gaussian and calculated the
improvement of the fit with a two-Gaussian model. The distribution of
probability ratios $\Delta P$ is shown in Figure~\ref{fig:frealtest1}.
The value we obtained for our CVnI data is shown as the single dot in
the Figure (upper panel). As this point coincides with the peak
obtained by fitting two Gaussians to artificial data consisting of a
single population, we conclude that we do not see evidence for a
second population in our data. The panel on the bottom of
figure~\ref{fig:frealtest1} shows the equivalent test for the
~\cite{Ibata2006} data \citep[where we have taken the data for their
26 stars with S/N $>15$ as listed in Table~2 of ][]{Martin2007}. In
this case the improvement is larger than would be expected to arise by
chance in a single-Gaussian data set. We note that a similar result is
obtained when the two data sets are combined. Therefore we conclude
that there is evidence of a second population containing $40$ per cent
of the total number of stars, in the \cite{Ibata2006} data set, at
almost the $3\sigma$ confidence level. We find that the dispersions of
these populations are $0.6$\,km\,s$^{-1}$ and $13.6$\,km\,s$^{-1}$,
respectively. Although these values are similar to those found by
\cite{Ibata2006}, we note that the populations we have identified may
be different to those in that paper, as in that case the separation of the
populations included an explicit velocity cut.
 
\subsection{Detectability}
\label{sec:det}
\begin{figure*}
 \begin{minipage}[hl]{0.3\linewidth}
  \centering 
  \includegraphics[scale=0.3]{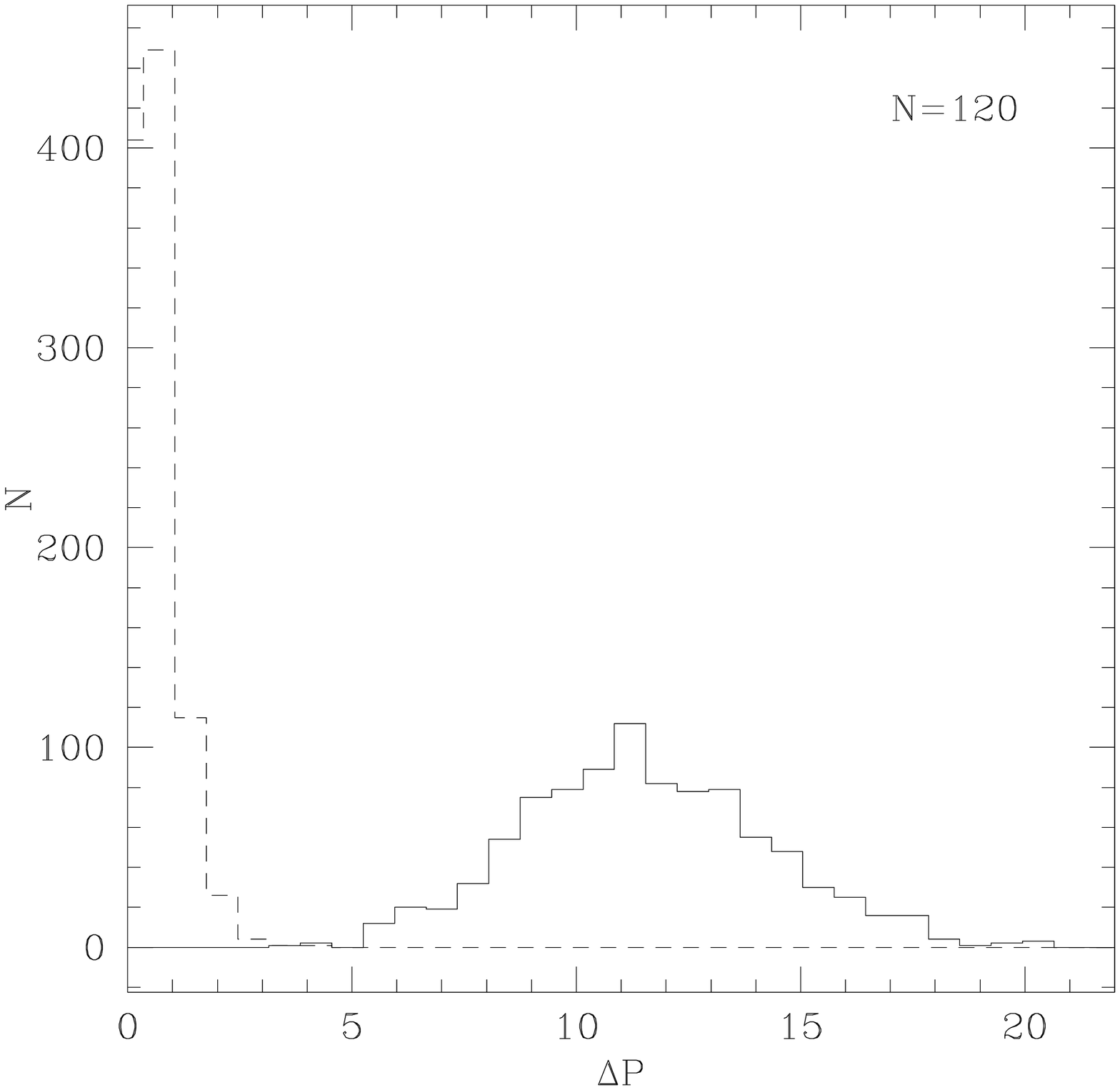}
\end{minipage}
\hspace{0.5cm}
   \begin{minipage}[h]{0.3\linewidth}
        \centering
  \includegraphics[scale=0.3]{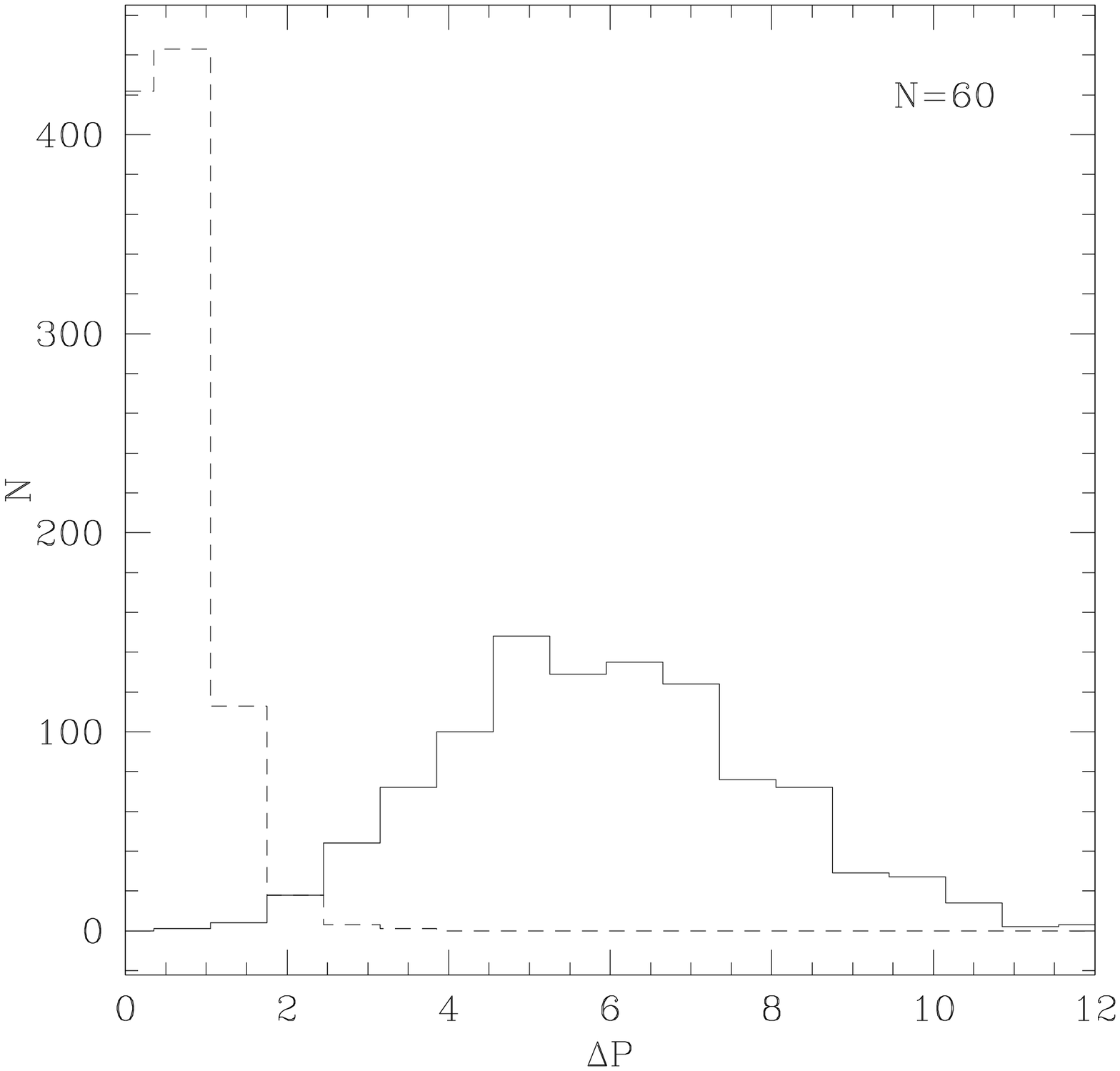}
\end{minipage}
\hspace{0.5cm}
    \begin{minipage}[h]{0.3\linewidth}
        \centering
  \includegraphics[scale=0.3]{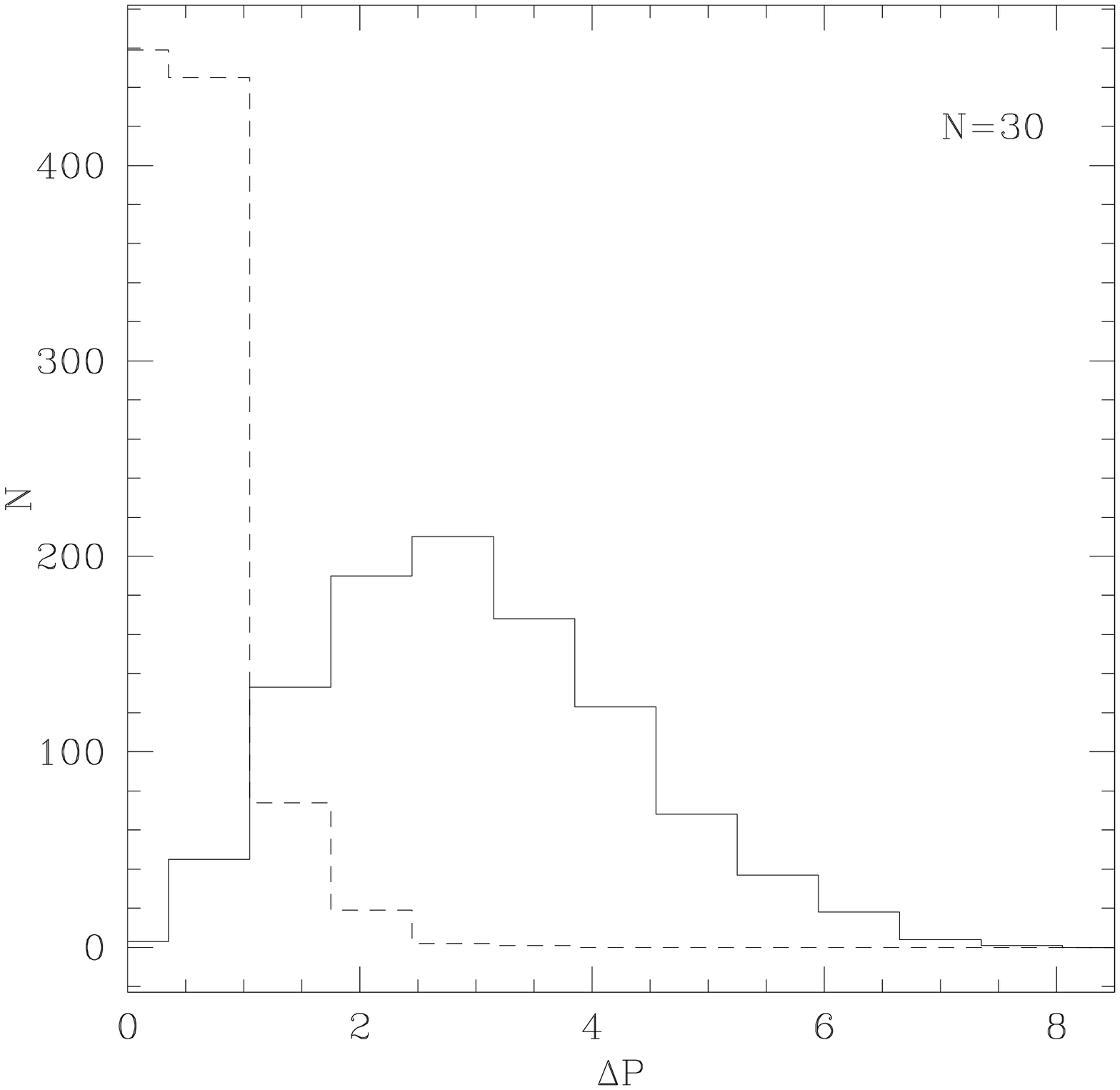}
\end{minipage}\caption{Histograms illustrating the detectability of 
sub-populations as a function of sample size. The two histograms in
each panel show the distribution of likelihood differences $\Delta P =
\log P_2-\log P_1$ between a single-Gaussian fit ($P_1$) and a two-Gaussian fit
($P_2$) for data sets having either a single population (dashed
histogram) or two equal-sized populations (solid histogram).  The
velocity error is constant: $dv=2$\,km\,s$^{-1}$. The total number of
stars is 120 in the left panel, 60 in the middle panel and 30 in the
right panel. As expected, the histograms merge together as the number
of stars decreases and the detection of a sub-population becomes more
difficult.}
\label{fig:fnumber}
\end{figure*}

Having considered our CVnI data set, in this section we investigate
the more general question of when sub-populations can be reliably
detected in small kinematic data sets. A limitation of our study is
that, for simplicity, we are working entirely with Gaussian
populations. However, our results will be conservative in the sense
that mixtures of non-Gaussian populations are likely to be more
difficult to disentangle.

The detectability of a sub-population depends on i) the total number of
stars in the data set; ii) the fraction of stars in the sub-population;
iii) the difference in velocity dispersion between the populations;
iv) the observational errors on the velocities. We investigate the
importance of each of these in turn. 
  
The total number of the stars is crucial for the detection of multiple
populations.  Figure~\ref{fig:fnumber} shows three tests done with
data sets of $N=120$,~$60$ and $30$ stars.  In each panel a comparison
is made between data sets that have either a single population or two
populations containing equal numbers ($N/2$) of stars. Motivated by
the case of CVnI, we consider data sets having two sub-populations
with $\sigma=13$\,km\,s$^{-1}$ and $\sigma=1$\,km\,s$^{-1}$. The two
populations are thus clearly distinct and we are thus isolating the
effect of the sample size in the result.  In Figure~\ref{fig:fnumber},
we plot the improvement in probability obtained using a two-Gaussian
fit to the single and double populations as dashed and solid
histograms, respectively. We determine the $1\sigma$ and $3\sigma$
range of the distribution of values obtained from the single
population (control) sample , and define a ($1\sigma$) $3\sigma$
detection of a sub-population to be one in which $\Delta P$ is larger
than the $1\sigma$ ($3\sigma$) limits of the control sample. Although
the difference between the dispersions is large, as we reduce the
sample size, the significance of the detection of multiple populations
decreases, as would be expected. Nevertheless, even for $N=30$ stars,
in 34.8 per cent (97.5 per cent) of cases, the subpopulation is
detected at the $3\sigma$ ($1\sigma$) confidence level.

\begin{figure}
  \includegraphics[width=0.5\textwidth]{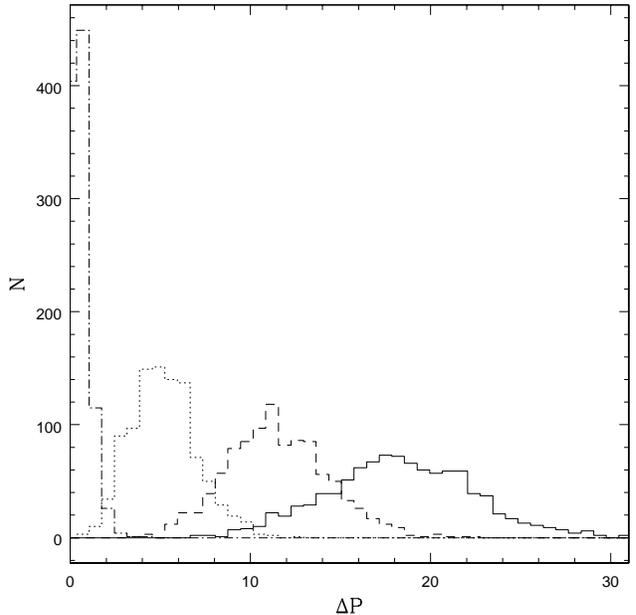}
\caption{Histograms illustrating the effect of the relative
sizes of the populations. As in previous figures, the histograms show
the distribution of likelihood differences $\Delta P = \log P_2-\log
P_1$ between a single-Gaussian fit ($P_1$) and a two-Gaussian fit
($P_2$) to the data sets. The histograms are plotted for a single
population (dot-dashed line), and for double populations including a
cold sub-population consisting of 30 per cent (dotted), 50 per cent
(dashed line) and 70 per cent (solid line) of the total sample of 120
stars. See text for a discussion.}
\label{fig:ffraction}
\end{figure}

The next parameter that we study is the fractional size of the
sub-populations.  This is important for the CVnI populations, since it
is possible that a cold population in the centre could have been
missed in our sample if it contained a smaller number of stars. Our
preliminary tests showed that a cold population could not be detected
even in a large sample if it only made up $\sim 0.1$ of the total
number of stars. Figure~\ref{fig:ffraction} shows results for cold
populations with fractional sizes $f=0.3$ , $f=0.5$ and $f=0.7$. In
this test, the dispersions of the sub-populations are
$13$\,km\,s$^{-1}$ and $1$\,km\,s$^{-1}$ and the velocity error is
$2$\,km\,s$^{-1}$. For $120$ stars, a $3\sigma$ detection was made for
all the samples with a cold population of fractional size $f=0.5$ and
$f=0.7$.  We found (see Figure~\ref{fig:ffraction}) that when the cold
population has a smaller fractional size in the sample i.e. $f=0.3$,
it was detected in 75.1 per cent (99.8 per cent) of cases at the
$3\sigma$ ($1\sigma$) level. It is thus easier to detect a
sub-population if its dispersion is larger than that of the main
population, rather than a cold sub-population.

\begin{figure}
  \includegraphics[scale=0.4]{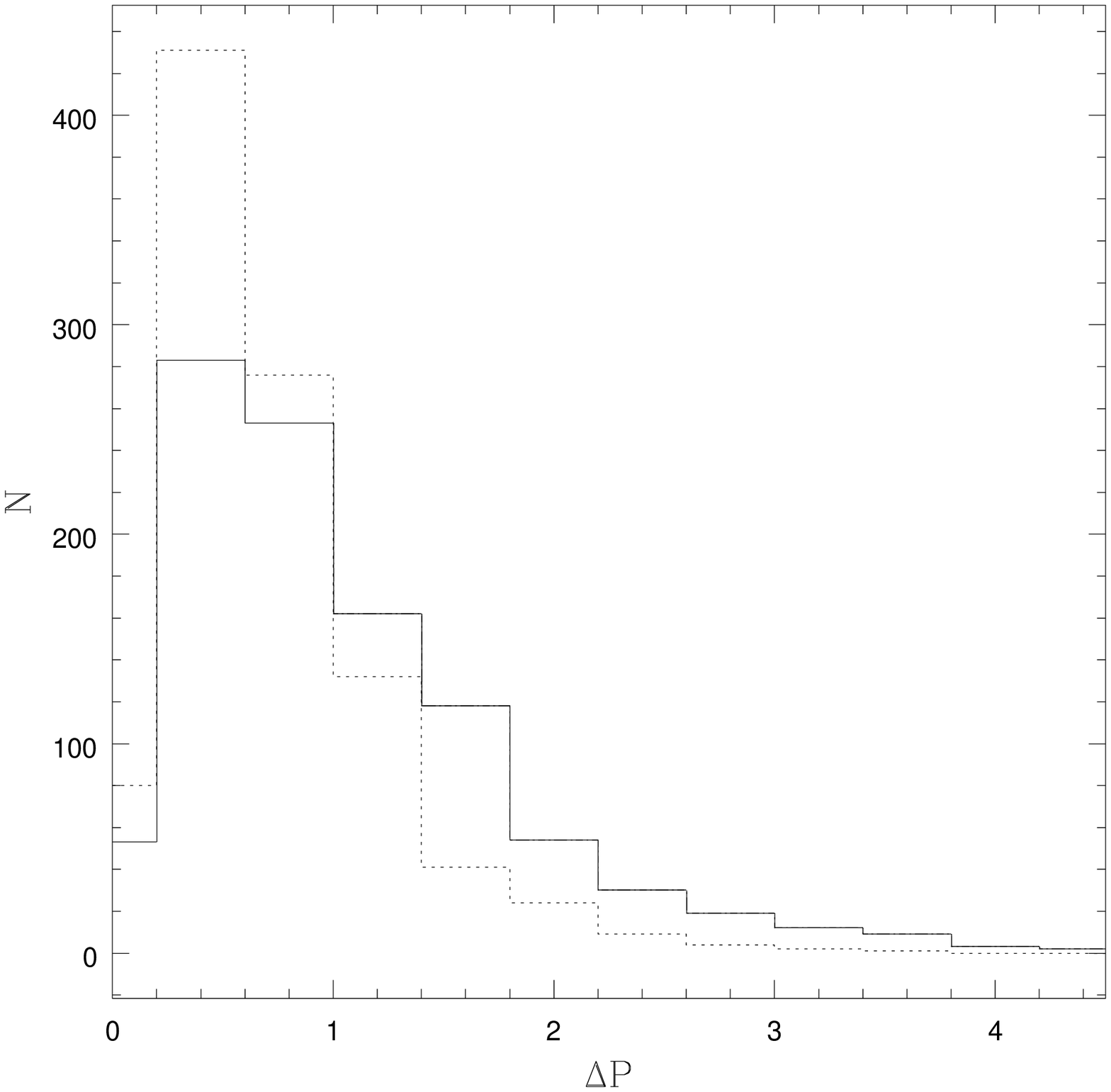}
  \includegraphics[scale=0.4]{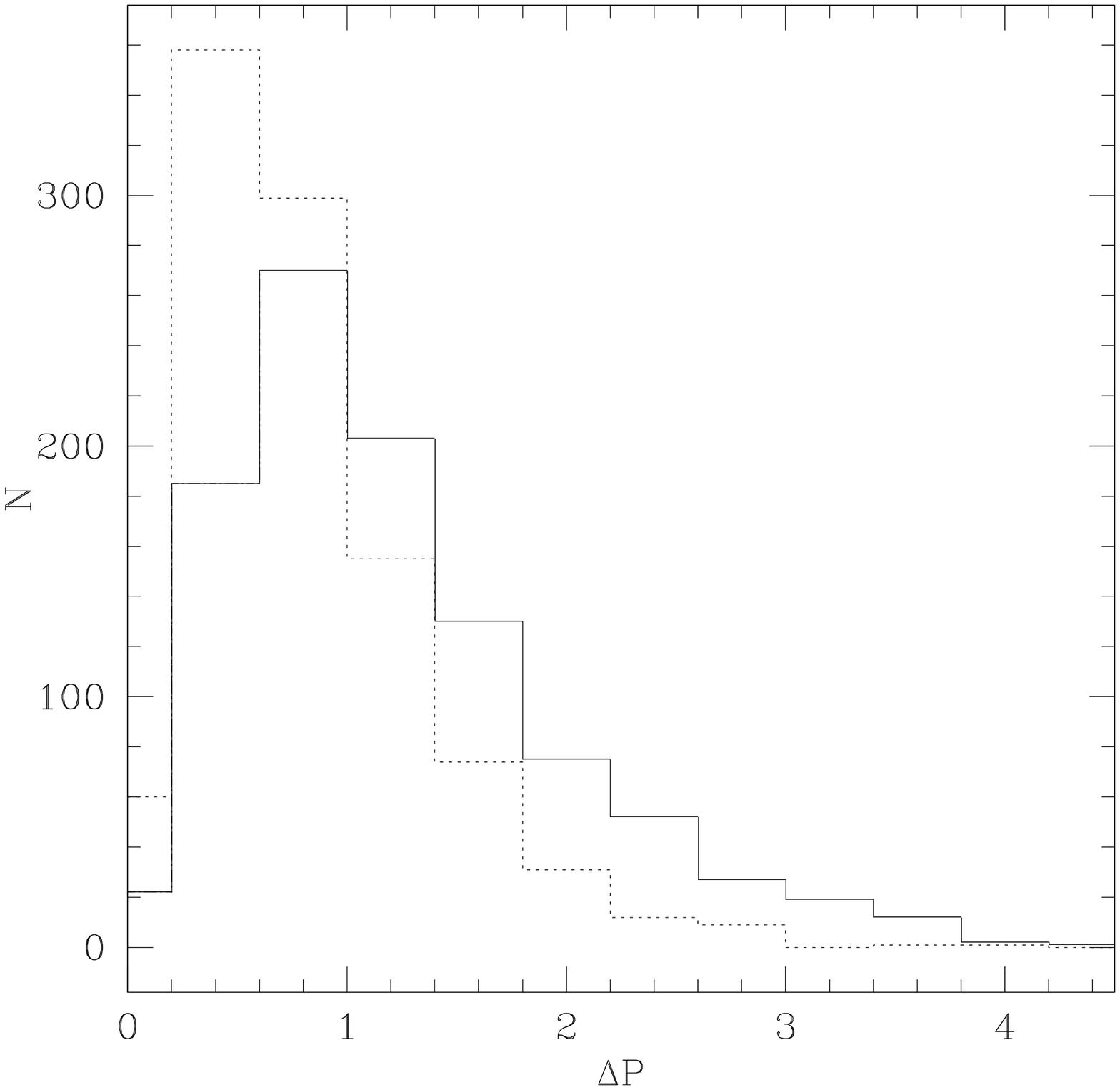}
\caption{Histograms illustrating the effect of velocity errors on the detection of
sub-populations with similar velocity dispersions. The total number of
stars is 120 and the sub-populations contain equal numbers of
stars. In each case, the double-populated sample with $\sigma_1 =
7$\,km\,s$^{-1}$ and $\sigma_2 = 4$\,km\,s$^{-1}$ is shown by the
solid-line histogram and compared to a single population system with
$\sigma= 7$\,km\,s$^{-1}$, shown as a dotted histogram.  In the top
panel, the velocity errors for both histograms are $dv =
2$\,km\,s$^{-1}$. As this error is relatively large compared to the
difference in the dispersion, in the bottom panel we repeat the same
experiment with $dv = 1$\,km\,s$^{-1}$.}
\label{fig:fdv}
\end{figure}

We next consider the impact of velocity errors on our ability to
detect multiple populations with similar velocity dispersions. The
sub-populations in this case have $\sigma_1=7$\,km\,s$^{-1}$ and
$\sigma_2=4$\,km\,s$^{-1}$. As Figure~\ref{fig:fdv} shows, decreasing
the errors from $dv=2$\,km\,s$^{-1}$ to $dv=1$\,km\,s$^{-1}$ gives
rise to a small change in the distribution of $\Delta P$
values. However, this does not lead to a significant increase in the
probability of detecting the multiple populations. We therefore
conclude that velocity errors at the $1-2$\,km\,s$^{-1}$ (similar to
the CVnI data) do not affect our ability to identify sub-populations.

Finally, to see the effect of the difference between the velocity
dispersions of the populations we investigate samples in which the
main population has a dispersion of $15$\,km\,s$^{-1}$ while the cold
sub-populations have dispersions ranging from $\sigma=1$\,km\,s$^{-1}$
to $14$\,km\,s$^{-1}$. We consider two sample sizes, with a total
number of either $120$ or $60$ stars. We find that even for a
relatively large sample of stars ($N=120$), a $3\sigma$ detection is
possible for all the samples only when
$\sigma_{1}/(\sigma_{1}-\sigma_{2})\leq1.1$, i.e. when the velocity
dispersions of the individual populations are
$\sigma_{1}=15$\,km\,s$^{-1}$ and $\sigma_{2}=1$\,km\,s$^{-1}$. A
$1\sigma$ level detection is possible for all 1000 samples for
$\sigma_{1}/(\sigma_{1}-\sigma_{2})\leq1.3$, in which case the
sub-populations' dispersions are $\sigma_{1}=15$\,km\,s$^{-1}$,
$\sigma_{2}=3$\,km\,s$^{-1}$. However populations with
$\sigma_{1}/(\sigma_{1}-\sigma_{2})\leq1.4$ and
$\sigma_{1}/(\sigma_{1}-\sigma_{2})\leq1.9$ can be detected at
$3\sigma$ and $1\sigma$ levels for 68 per cent of the 1000
samples. For a smaller sample ($N=60$), a $3\sigma$ detection for all
samples is not possible for even the largest ratios of of
$\sigma_{1}/\sigma_{2}$. In this case $3\sigma$ and $1\sigma$
detections for 68 per cent of the samples require
$\sigma_{1}/(\sigma_{1}-\sigma_{2})\leq1.3$ and
$\sigma_{1}/(\sigma_{1}-\sigma_{2})\leq1.7$ respectively.  We note
that the claimed CVnI populations in ~\cite{Ibata2006} have an even
more extreme dispersion difference
$\sigma_{1}/(\sigma_{1}-\sigma_{2})=1.04$. A Monte Carlo experiment
with 26 stars and this dispersion ratio shows that in this case
populations can be identified with $3\sigma$ confidence in 90.5 per
cent of the samples. Table~\ref{tab:tdet} summarises our results for
the full range of dispersions ratios we have considered.

∑

 \begin{table}
\begin{center}
\begin{tabular} {l c c r}
\hline
$\sigma_{1}/(\sigma_{1}-\sigma_{2})$ & N &  $N> \sigma$  & $ \rm N > 3 \sigma$ \\\hline
                         & $120$ stars &$1000$& $1000$  \\ [-1ex]
\raisebox{1.5ex}{$1.1$}  & $60$ stars &$1000$ & $984$  \\ [1ex]
                         & $120$ stars &$1000$& $999$ \\  [-1ex]
\raisebox{1.5ex}{$1.2$}  & $60$ stars &$999$ & $895$  \\ [1ex]
                         & $120$ stars &$1000$& $963$   \\  [-1ex]
\raisebox{1.5ex}{$1.3$}  &  60 stars   &$989$ & $685$ \\ [1ex]
                         & $120$ stars &$997$ & $782$   \\  [-1ex]
\raisebox{1.5ex}{$1.4$}  &  60 stars  & $946$ & $397$ \\ [1ex]
                         & $120$ stars &$966$ & $480$   \\  [-1ex]
\raisebox{1.5ex}{$1.5$}  &  60 stars &  $851$ & $189$ \\ [1ex]
                         & $120$ stars &$874$ & $203$   \\  [-1ex]
\raisebox{1.5ex}{$1.7$}  &  60 stars &  $711$ & $82$ \\ [1ex]
                         & $120$ stars &$738$ & $72$   \\  [-1ex]
\raisebox{1.5ex}{$1.9$}  &  60stars &   $572$ & $31$ \\ [1ex]
                         & $120$ stars &$562$ & $28$   \\  [-1ex]
\raisebox{1.5ex}{$2.1$}  & 60 stars &   $451$ & $12$ \\ [1ex]
                         & $120$ stars &$424$ & $5$   \\  [-1ex]
\raisebox{1.5ex}{$2.5$}  & 60 stars &   $347$ & $4$ \\ [1ex]
                         & $120$ stars &$313$ & $4$   \\  [-1ex]
\raisebox{1.5ex}{$3.$}   & 60 stars &   $276$ & $2$ \\ [1ex]
                         & $120$ stars &$227$ & $1$   \\  [-1ex]
\raisebox{1.5ex}{$3.8$}  & 60 stars &   $228$ & $2$ \\ [1ex]
                         & $120$ stars &$193$ & $0$   \\  [-1ex]
\raisebox{1.5ex}{$5$}    &  60 stars &  $195$ & $1$ \\ [1ex]
                         & $120$ stars &$160$ & $0$   \\  [-1ex]
\raisebox{1.5ex}{$7.5.$} &  60 stars &  $174$ & $1$ \\ [1ex]
                         & $120$ stars &$145$ & $0$   \\  [-1ex]
\raisebox{1.5ex}{$15$}   & 60 stars &   $154$ & $1$ \\ [1ex]
\end{tabular}
\caption{Confidence limits for the detection of sub-populations with
different kinematics. Columns are: (1) Ratio of main velocity
dispersion $\sigma_1$ to the difference between the populations
$\sigma_1-\sigma_2$; (2) total number of stars in the data set; (3)
Number of two-population samples for which $\Delta P$ is greater than
the $1\sigma$ upper limit of $\Delta P$ obtained from
single-population samples; (4) Number of two-population samples for
which $\Delta P$ is greater than the $3\sigma$ upper limit of $\Delta
P$ obtained from single-population samples. We compare populations of
60 and 120 stars containing two sub-populations. In each case
$\sigma_1=15$\,km\,s$^{-1}$ while $\sigma_2$ lies in the range
$1$\,km\,s$^{-1}$ to $14$\,km\,s$^{-1}$. Each dispersion ratio has been
tested for 1000 data sets.}
\label{tab:tdet}
\end{center}
\end{table}

\section{Conclusions}
\label{sec:conc}

In this paper, we have presented a new data set of velocities and
metallicities for the Canes Venatici I (CVnI) dSph based on spectra
taken with the GMOS-North spectrograph. A maximum likelihood fit to
the velocity distribution yields a mean velocity of $v=
25.8\pm0.3$\,km\,s$^{-1}$ and a dispersion of $\sigma
=7.9^{+1.3}_{-1.1}$\,km\,s$^{-1}$. Assuming a constant, isotropic
velocity dispersion and a Plummer profile for the mass distribution,
we find a mass of $4.4^{+1.6}_{-1.1}\times10^7 M_\odot$ in the volume
where our tracer stars are located. Although this value is larger than
the value $2.7\pm 0.4\times 10^7 M_\odot$ calculated by
~\cite{Simon2007},this is most likely due to the assumptions made for
our models and the distribution of our particular subsets of stars.

One of the original aims of our study was to investigate the claimed
multiple stellar populations in CVnI. As we discussed above, the two
previous studies by ~\cite{Ibata2006} and ~\cite{Simon2007} did not
agree on the existence of a cold sub-population in CVnI. The two
populations found in the former study were puzzling as they led to two
different mass estimates. The authors suggested that this might
indicate that the system had recently accreted a younger population
and was not yet in equilibrium.

In this paper we looked for evidence of multiple populations in our
data under the assumption that each population was Gaussian. Based on
this analysis, we concluded that there was no reason to suspect the
presence of a second population in our data. We also applied our
analysis to the \cite{Ibata2006} data where we found evidence of a
statistically significant sub-population with a dispersion of $\sigma
=0.6$\,km\,s$^{-1}$ (compared to $\sigma =13.6$\,km\,s$^{-1}$ for
the main population). 

Our analysis suggests that there is a qualitative difference between
our data and those of \cite{Ibata2006}. Although further data would be
necessary to resolve this issue, we note that the spatial
distributions of these two data sets are different, which could
potentially account for the differences in the detected populations.
However, our central field is centred close to the blue/young star population which
\cite{Martin2008a} find in their photometry from the Large Binocular
Telescope, and which they identify with the cold population of
\cite{Ibata2006}. The exact fraction of stars in each population found by
\cite{Martin2008a} is currently unclear, however, and so it is possible that
we have not picked up any stars associated with the cold population.

We have also carried out a study of the detectability of
sub-populations in small kinematic data sets.  Under the assumption of
Gaussian populations, we studied the effects of four parameters. We
obtained confidence limits for the detection of sub-populations in
samples with different numbers of stars, different population ratios
and velocity dispersions. We found that reasonable errors on the
observed velocities do not affect the detectability of the
sub-populations. For a given sample size, our ability to detect two
populations increased as the ratio of their dispersions
$\sigma_{1}/\sigma_{2}$ increased. However, even for large
$\sigma_{1}/\sigma_{2}$ and equal population size, a sample of 30
stars yielded a $3\sigma$ detection in only $\sim35$ per cent of
cases. As expected, for larger sample sizes, this detection rate was
significantly higher. We also showed that a cold population needs to
constitute a larger fraction of the total sample than is required to
detect a hot sub-population. This suggests that the robust detection
of the sub-populations associated with any surviving sub-haloes within
a dSph would require samples of many hundreds of velocities. In this
case, localised substructures could be detected by windowing the data,
provided that a window whose spatial size coincided with plausible
sub-halo scales would contain a sample of at least 100 stars. As such
data sets are now becoming available for many of the larger dSphs,
this test may soon be feasible. We note that the claim of multiple
global populations in Sculptor~\citep[]{Tolstoy2004} was based on a
large data set and is therefore still robust.

Finally, we note that all our significance tests were based on the
assumption of Gaussian populations, which was the case for all our
Monte Carlo samples. However, for real data, the true distributions
will not be known, and are not necessarily well-approximated by
Gaussians. It is therefore difficult in a real case to assign a robust
statistical significance to a particular detection of a
sub-population.

As we have shown, for small data sets, many Monte Carlo realisations
do not yield significant detections of the sub-populations. In the
absence of a robust estimate of the confidence level of a particular
detection, or additional, independent evidence of the presence of
multiple populations, we conclude that one should exercise great
caution in decomposing data sets of fewer than $100$ stars into
multiple populations.

\section*{Acknowledgments}
UU acknowledges funding from the European Commission under the Marie
Curie Host Fellowship for Early Stage Research Training SPARTAN,
Contract No MEST-CT-2004-007512, University of Leicester, UK. MIW
acknowledges support from a Royal Society University Research
Fellowship. T.C.B. acknowledges partial funding of this work
from grants AST 07-07776 and PHY 02-16783: Physics Frontiers Center / Joint
Institute for Nuclear Astrophysics (JINA), awarded by the U.S. National
Science Foundation.

Based on observations obtained at the Gemini Observatory, which is operated by
the Association of Universities for Research in Astronomy (AURA) under a
cooperative agreement with the NSF on behalf of the Gemini partnership: the
National Science Foundation (United States), the Science and Technology
Facilities Council (United Kingdom), the National Research Council (Canada),
CONICYT (Chile), the Australian Research Council (Australia), CNPq (Brazil)
and CONICET (Argentina). Program ID: GN-2007A-Q-66.

\end{document}